\def\year{2019}\relax
\documentclass[letterpaper]{article} 
\usepackage{aaai19}  
\usepackage{times}  
\usepackage{helvet}  
\usepackage{courier}  
\usepackage{url}  
\usepackage{graphicx}  
\usepackage{booktabs} 
\usepackage{amsmath,amssymb} 
\usepackage{color}

\usepackage{multirow}
\usepackage{algorithm}
\usepackage{algorithmicx}
\usepackage{algpseudocode}
\usepackage{subfig}
\usepackage{comment}
\usepackage{float}

\begin{document}
%
\title{Distribution Discrepancy Maximization for Image Privacy Preserving}
\author{Sen Liu, Jianxin Lin, Zhibo Chen\\
Intelligent Media Computing Lab, University of Science and Technology of China\\
elsen@iat.ustc.edu.cn, linjx@mail.ustc.edu.cn, chenzhibo@ustc.edu.cn
}
\maketitle
\newcommand{\citelp}[1]{\citeauthor{#1}~\shortcite{#1}}
\begin{abstract}

With the rapid increase in online photo sharing activities, image obfuscation algorithms become particularly important for protecting the sensitive information in the shared photos. However, existing image obfuscation methods based on hand-crafted principles are challenged by the  dramatic development of deep learning techniques. To address this problem, we propose to maximize the distribution discrepancy between the original image domain and the encrypted image domain. Accordingly, we  introduce a collaborative training scheme: a discriminator $D$ is trained to discriminate the reconstructed image from the encrypted image, and an encryption model $G_e$ is required to generate these two kinds of images to  maximize the recognition rate of $D$, leading to the same training objective for both $D$ and $G_e$. We theoretically prove that such a training scheme maximizes two distributions' discrepancy. Compared with commonly-used image obfuscation methods, our model can produce satisfactory defense against the attack of deep recognition models indicated by significant accuracy decreases on FaceScrub, Casia-WebFace and LFW datasets.

\end{abstract}

\section{Introduction}

With the popularization of mobile devices carrying high resolution cameras and the explosiveness of social networks, there has been significant increase in personal photo sharing online. While photo sharing has become part of our daily life, privacy concerns are raised in photo leakage and photo recognition by unauthorized users or algorithms. Image obfuscation algorithms meanwhile are proposed to address the challenge of image privacy leakage by obfuscating the sensitive area in the images, such as faces, logos or objects. For example, Pixelation and Blurring are the two most classic approaches that suppress recognizable sensitive features while keeping everything else integral. However, encrypting photos by destroying information can also result in difficult restoration unless original ones are backed up, which requires much more resources for network transmission and cloud storage.

On the other hand, \citelp{ra2013p3} proposed a reconstructable obfuscation algorithm, named P3, which decomposes image's JPEG coefficients into public part and privacy part. However, these traditional approaches omit to encrypt faces or other sensitive objects to be unrecognizable by well-defined principles. In addition, simple JPEG coefficients decomposition method fails to resist attack from deep learning metrics which have shown dramatic improvement on many high-level computer vision problems, such as face recognition \cite{parkhi2015deep}, image recognition \cite{karen2014very} and so on. Compared with traditional image attack algorithms, deep learning based approaches have stronger ability to represent semantic structures \cite{zeiler2014visualizing}, and are more powerful to handle high-dimensional image data. Some studies \cite{oh2016faceless,mcpherson2016defeating} have shown that deep recognition models could extract privacy information from human unrecognizable data encrypted by some image obfuscation techniques.

\begin{figure}[t]
\begin{center}
   \includegraphics[width=1\linewidth]{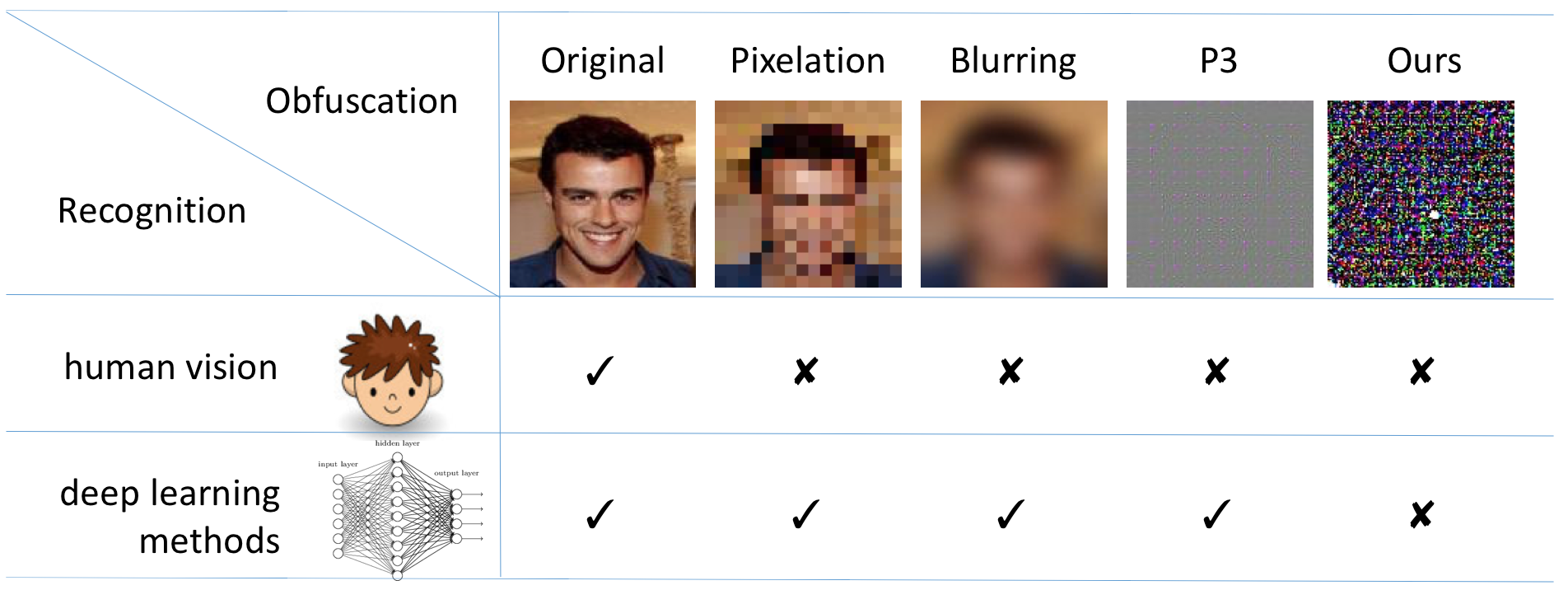}
\end{center}
   \caption{Given an input image, our goal is to make obfuscation for protecting its privacy information. Our approach, which is based on distribution discrepancy maximization, has better capability to defend attack from human visual recognition and deep recognition models.}
\label{fig:intro}
\end{figure}

To address the challenge from the deep learning, we propose to maximize the distribution discrepancy between the original image domain and the encrypted image domain. First, given an original input image, an encryption model encodes the image into deep features which are decomposed into two parts, called public feature and privacy feature. Second, the encryption model is required to generate reconstructed image (from public feature and privacy feature) and encrypted image (from public feature) to maximize the recognition rate of a discriminator. Meanwhile, the discriminator is trained to discriminate these two kinds of images. So the encryption model and discriminator keep the same training objective, which is called as ``collaborative training scheme''. We theoretically prove that such training scheme maximizes the two distributions' discrepancy. Compared with Generative Adversarial Nets (GAN) \cite{goodfellow2014generative}, where a generator is trained to maximize the probability of discriminator making a mistake, the proposed collaborative training scheme changes the relationship between generator and discriminator, where both networks work as collaboration instead of competition. Third, we minimize the pixel-wise and perceptual difference between the input and the reconstructed image, which ensures that reconstructed images and the original images share the same image domain. As a result, our model learns to squeeze the privacy information into the privacy feature and produce remarkable distribution discrepancy between the encrypted images and the input images. We also empirically demonstrate that distribution discrepancy maximization can effectively protect the sensitive information of images from the attack of deep recognition models and human visual recognition, while a small amount of information is required to be encrypted.

The contributions of this paper are summarized as follows:

\begin{itemize}
\item We propose to maximize the distribution discrepancy for image privacy preserving. Our proposal is effective to defend recognition from human and deep learning based methods.
\item We introduce a collaborative training scheme which is theoretically proved to maximize the distribution discrepancy between two image domains.
\end{itemize}

The remaining parts are organized as follows. We introduce related work in Section \ref{related work} and present the details of our method in Section \ref{framework}. The theoretical analysis of collaborative training scheme is given in Section \ref{theo_ana}. The implement details are described in Section \ref{implement}. Then we report experimental results in Section \ref{experiment} and conclude in Section \ref{conclusion}.
\section{Related Work}\label{related work}
\begin{figure*}[htbp]
\begin{center}
   \includegraphics[width=15cm]{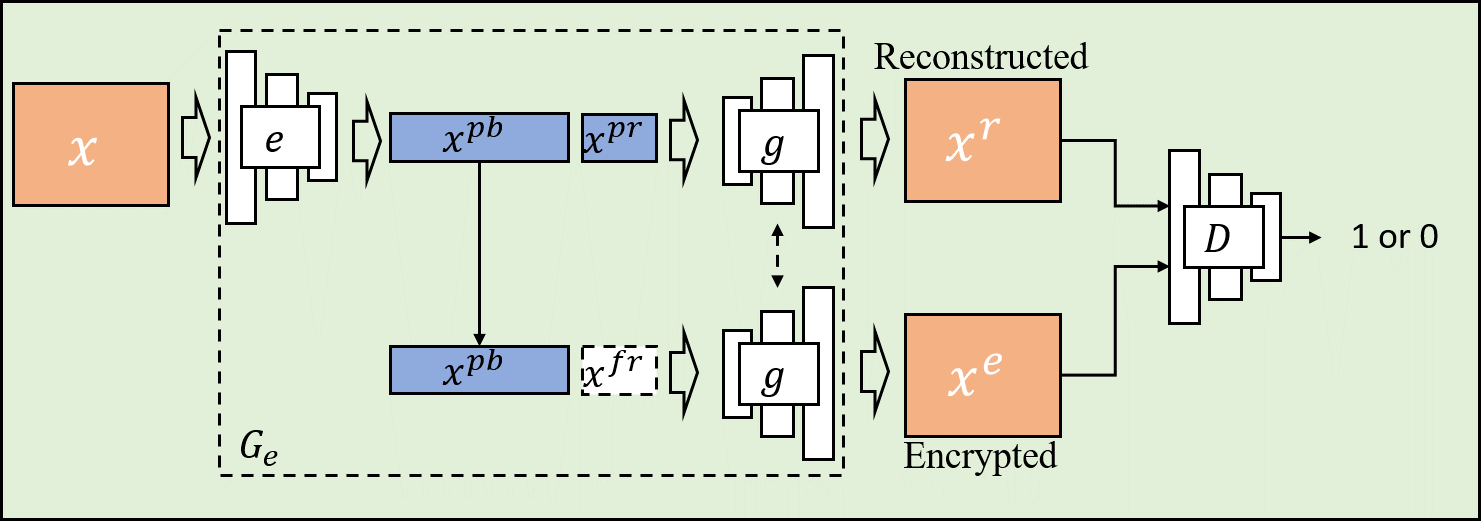}
\end{center}
   \caption{Architecture of encryption model and discriminator.}
\label{fig:arch2}
\end{figure*}
\subsection{Image Obfuscation Algorithms}
There are mainly three kinds of image obfuscation algorithms, including Blurring, Pixelation and P3. Both Blurring and Pixelation share similar encryption principle and have been widely applied to encrypt face images and sensitive objects. Pixelation first divides the image into square grids. Then the average pixel value of current grid is assigned to each entire grid, which is similar to an average pooling operation applied for images. \citelp{ryoo2017privacy-preserving} suggested that reliable face recognition from extreme low resolution scheme (with scale factor x20) will be difficult for both machines and humans. Blurring convolves the image with Gaussian kernel to obscure the detail of image, which is similar to a convolutional operation with fixed values. In $2012$ and $2016$, YouTube utilizes Blurring as a way to protect video privacy and introduces to automatically blur faces in \cite{youtube2012face} and objects in \cite{youtube2016face}. However, Blurring and Pixelation do not wipe all image or video information off, and have been demonstrated that they may degrade the visual recognition but are insufficient for privacy protection \cite{gopalan2012blur}.

\citelp{ra2013p3} proposed a privacy-preserving photo encoding algorithm called P3. They decomposed the JPEG image into a public image and a secret image. The decomposition operation is based on assumption that DC coefficients and AC coefficients whose values are larger than a threshold carry the most information about the image, which should be encrypted and extracted in the secret image. Public image is excluded sensitive information and can be uploaded it to public servers. Our work shares the similar philosophy with P3. The difference is that the decomposing operation is conducted on deep features instead of JPEG coefficients. In addition, we propose to learn to decompose privacy part and public part using a collaborative training scheme instead of manually designed technique, which is more robust under the attack of deep recognition models.

\subsection{Attack Methods for Decrypting the Encrypted Images}
Traditional visual recognition works usually extract hand-crafted features, including SIFT \cite{lowe2004distinctive}, HOG \cite{dalal2005histograms}, and feed them to classifiers. For example, \citelp{viola2004robust} proposed a fast face detection method based on low-level features aggregation and cascade classifiers combination. A classic face recognition algorithms take a face images as input and match it with face eigenvectors of a face dataset, such as Eigenface \cite{turk1991eigenfaces}. With the dramatic development of deep learning, deep learning based methods have shown STOA performance on many recognition tasks. The capability of deep learning also has been utilized to discover the relevant features or relationship in encrypted images. Works in \cite{oh2016faceless,mcpherson2016defeating} employed deep recognition models to successfully defeat common image obfuscation algorithms such as Pixelation, Blurring and P3. In addition, DNN also shows outstanding ability to restore images from severe noise and low-resolution, such as methods in \cite{mao2016image,yu2016ultra}, which can also be used as a pre-processing step for decrypting. The experimental results of these methods reveal that well-designed deep learning methods can still recognize the sensitive information of encrypted images.

\subsection{Adversarial Attack on Deep Learning}

As deep learning achieves significant performance in visual recognition and detection tasks, several works also focused on attacking the powerful deep models. These methods tried to build a modified version of a clean image that is intentionally perturbed to confuse a deep neural networks, which misclassifies the image with high probability \cite{goodfellow2015explaining,mopuri-bmvc-2017,raval2017protecting,oh2017adversarial,Moosavi-Dezfooli_2017_CVPR}. It's worth to note that this adversarial perturbation changes the visual information of images slightly, which is hardly noticed by human. The essential difference between adversarial perturbation methods and ours is the human recognizability of images after encryption. The goal of our work is to defend the attack from deep recognition methods as well as encrypting images being unrecognizable by human. Experiment results in Section \ref{Comparison Results} also show that our work achieves better performance on the attack of deep recognition models.


\section{Our Method}\label{framework}
\subsection{Problem Formulation}\label{sec:problem}

We first formulate the decomposition and reconstruction process to clarify image privacy preserving problem. Given one image input $x$. Following the basic assumption of our paper, $x$ can be represented by two kinds of features as $x=x^{pb} \oplus x^{pr}$, where $x^{pb}$ is public feature, $x^{pr}$ is privacy feature and $\oplus$ is the operator that can merge the two kinds of features into a complete image. Then the problem of privacy preserving problem is as follows: taking an image $x$ as input and outputting a reconstructed image $x_{r}$ that is the estimation of the original input image, and encrypted image $x_{e}$ that is unrecognizable from original input image, i.e.,
\begin{equation}
\begin{aligned}
x_{r} = G_{x\rightarrow r}(x) = x^{pb} \oplus x^{pr},
\end{aligned}
\label{eq:our_prime_task_1}
\end{equation}
\begin{equation}
\begin{aligned}
x_{e} = G_{x\rightarrow e}(x) = x^{pb} \oplus x^{fr},
\end{aligned}
\label{eq:our_prime_task_2}
\end{equation}
where $G_{x\rightarrow r}$ denotes the reconstruction function, $G_{x\rightarrow e}$ denotes the encryption function, and $x^{fr}$ is the faked privacy feature. $x^{fr}$ is set to $x^{pr}+n$, $n$ is AWGN noise (standard deviation = 1), which requires $x_{e}$ to be different from $x_{r}$ unless using the correct privacy feature. An implicit assumption of our problem formulation is that the public part is exposed during transportation while the privacy part is encrypted and transported as paradigm suggested in \cite{deng2004secure,mink2006high}. So privacy feature should include the key information and be as little as possible for transmission efficiency.
\subsection{Network Architecture}\label{sec:arch}
Our network architecture is shown in Figure \ref{fig:arch2}, which mainly includes one AutoEncoder-like encryption model $G_e$ and one discriminator $D$. We first feeds the original input image $x$ into the encoder network of $G_e$ to extract deep features. Then the deep features are decomposed into two part, public part $x^{pb}$ and privacy part $x^{pr}$. In particular, given one image input $x$, we have:
\begin{equation}
\begin{aligned}
(x^{pb},x^{pr}) = e(x),
\end{aligned}
\label{eq:encoder}
\end{equation}
where $e$ is the encoder network of $G_e$. We generate reconstructed image $x_{r}$ and encrypted image $x_{e}$ as following:
\begin{equation}
\begin{aligned}
x_{r} = g(x^{pb},x^{pr});\quad x_{e} = g(x^{pb},x^{fr}),
\end{aligned}
\label{eq:decoder}
\end{equation}
where $g$ is the decoder network of $G_e$. Currently, there is no obvious difference between public feature and privacy feature. In the remaining sections, we will present the overall training process that differentiates the two kinds of features.

\subsection{Collaborative Training Scheme}\label{sec:adv}
We introduce a collaborative training scheme for distribution discrepancy maximization between the original image domain and the encrypted image domain. We define a discriminator $D$ that is trained to discriminate the reconstructed image from the encrypted image, and a encryption model $G_e$ is required to generate encrypted image and reconstructed image maximizing recognition rate of $D$. The objective function of discriminator $D$ is given as:
\begin{equation}
\begin{aligned}
\begin{split}
\ell_{D} =& {\mathbb{E}_{x_r \sim {p_r}(x)}}[BCE(D(x_r),1)]\\&+{\mathbb{E}_{x_e \sim {p_e}(x)}}[BCE(D(x_e),0)],
\end{split}
\end{aligned}
\label{eq:d_loss}
\end{equation}
where $BCE$ is the binary cross entropy loss function. Accordingly, the collaborative objective function of the encryption model $G_e$ can be given as:
\begin{equation}
\begin{aligned}
\begin{split}
\ell_{G}^{\text{ad}} =& {\mathbb{E}_{x_r \sim {p_r}(x)}}[BCE(D(x_r),1)]\\&+{\mathbb{E}_{x_e \sim {p_e}(x)}}[BCE(D(x_e),0)].
\end{split}
\end{aligned}
\label{eq:g_loss}
\end{equation}
So the encryption model and discriminator have the same objective to optimize, which is called as ``collaborative training scheme''. In the Section \ref{theo_ana}, we will present a theoretical analysis of collaborative training scheme, which shows that such training process makes the distributions of the reconstructed images and the encrypted images be discrepant. Although here we maximize the distribution discrepancy between the encrypted samples and the reconstructed samples, we present experiment results that our model can also provide a well reconstruction of the original samples in Section \ref{Toy Experiment} and Section \ref{privacy_proportion}.

\subsection{Reconstruction}
Furthermore, image obfuscation model should also maintain the whole completeness of input image. Intuitively, this can be easily realized by constraining Mean Square Error (MSE) between reconstructed image $x_{r}$ and input image $x$ as following loss function:
\begin{equation}
\begin{aligned}
\ell_{G}^{\text{r,m}} = {\mathbb{E}_{x \sim {p}(x)}}[\Vert x_r - x\Vert_2^2].
\end{aligned}
\label{eq:recon_loss}
\end{equation}
We also leverage a perceptual loss \cite{johnson2016perceptual}, which depends on high-level features from a pre-trained loss network, such as VGG network \cite{karen2014very}, together with MSE loss for reconstruction constraint. Such multi-perceptual level constraint further enables the information reconstructed into $x_{r}$. The full objective function for reconstruction is:
\begin{equation}
\begin{aligned}
\ell_{G}^{\text{r}} =  \ell_{G}^{\text{r,m}} + \lambda{\mathbb{E}_{x \sim {p}(x)}}[\Vert \phi(x_r) - \phi(x)\Vert_2^2],
\end{aligned}
\label{eq:privacy_loss}
\end{equation}
where $\phi$ is one pre-trained network and $\lambda$ is set to $0.01$ for loss balance.

\subsection{Full Objective}\label{sec:loss}

Our full objective for encryption model $G_e$ is:
\begin{equation}
\begin{aligned}
\ell_{G}=\ell_{G}^{\text{ad}} +\ell_{G}^{\text{r}}.
\end{aligned}
\label{eq:full_loss_G}
\end{equation}
We summarize the training process in Algorithm~\ref{alg_1}. In Algorithm~\ref{alg_1}, the choice of optimizers $Opt(\cdot,\cdot)$ is quite flexible, whose two inputs are the parameters to be optimized and the corresponding gradients. We choose Adam~\cite{kingma2014adam} in our real implement. Besides, the $G_e$ and $D$ might refer to either the models themselves, or their parameters, depending on the context.

\begin{algorithm}
\caption{Collaborative training process}
\label{alg_1}
\begin{algorithmic}[1]
\Require Training images $\{x\}_{i=1}^{m}\subset\mathcal{D}$, batch size $K$, optimizer $Opt(\cdot,\cdot)$, weight parameters $(\lambda)$;
\State Randomly initialize $G_e$ and $D$.
\State Randomly sample a minibatch of images and prepare the batch training data $\mathcal{S}=\{x_{k}\}^K_{k=1}$.
\State For any data $x_{k}\in\mathcal{S}$, extract public feature and privacy feature by Eqn.(\ref{eq:encoder}), generate reconstructed image and encrypted image by Eqn.(\ref{eq:decoder});
\State Update the discriminator as follows:\newline
$D \leftarrow Opt(D, (1/K)\nabla_{D}\textstyle{\sum_{k=1}^{K}}\ell_{D}(x_{k}))$;
\State Update the encryption model as follows:\newline
$G_e \leftarrow Opt(G_e, (1/K)\nabla_{G_e}\textstyle{\sum_{k=1}^{K}}\ell_{G}(x_{k}))$;
\State Repeat step 2 to step 6 until convergence
\end{algorithmic}
\end{algorithm}

\section{Theoretical Analysis}\label{theo_ana}
\textbf{Proposition 1.} \textsl{Given a fixed $G_e$, the optimal discriminator $D$ is }
\begin{equation}
\begin{aligned}
D_{w,G}^{*}(x) = \frac{p_{r}(x)}{p_{r}(x)+p_{e}(x)}.
\end{aligned}
\label{eq:optimal_D}
\end{equation}
\textsl{Proof.} For a fixed encryption model $G_e$, the training criterion of $D$ is to minimize the loss function in Eqn.(\ref{eq:d_loss}). As shown in the Goodfellow et al. \cite{goodfellow2014generative}, it is easy to obtain the minimum at $\frac{p_{r}(x)}{p_{r}(x)+p_{e}(x)}$.

\textbf{Proposition 2.} \textsl{Under an optimal discriminator, the encryption model $G_e$ maximizes the Jensen-Shanon divergence.}

\textsl{Proof.} By inspecting Eqn.(\ref{eq:optimal_D}) into Eqn.(\ref{eq:g_loss}), we obtain:
\begin{equation}
\begin{aligned}
\begin{split}
\ell_{G,D^*}^{\text{ad}} =& {\mathbb{E}_{x_r \sim {p_r}(x)}}[BCE(D_{w,G}^{*}(x_r),1)]\\&+{\mathbb{E}_{x_e \sim {p_e}(x)}}[BCE(D_{w,G}^{*}(x_e),0)]\\
=&\text{log}(4)-KL\left(p_{r}\|\frac{p_r+p_e}{2}\right)\\&-KL\left(p_{e}\|\frac{p_r+p_e}{2}\right),
\end{split}
\end{aligned}
\label{eq:optimal_G}
\end{equation}
which $KL$ is the Kullback-Leibler divergence. Eqn.(\ref{eq:optimal_G}) can be rewritten in terms of the Jensen-Shannon divergence as:
\begin{equation}
\begin{aligned}
\begin{split}
\ell_{G,D^*}^{\text{ad}} = \text{log}(4) - 2JSD(p_{r}\|+p_{e}).
\end{split}
\end{aligned}
\label{eq:optimal_G_jsd}
\end{equation}
The Jensen-Shannon divergence between two distributions is always non-negative and achieves zero only when they are equal. Therefore, minimizing $\ell_{G,D^*}^{\text{ad}}$ can directly maximize the JSD between $p_{r}$ and $p_{e}$.
\section{Implement Details}\label{implement}

\subsection{Network Configuration}
For the encryption model, the encoder contains four stride-2 convolution layers, three residual blocks between two convolution layers, and the encoder contains four stride-2 deconvolution layers, three residual blocks between two deconvolution layers. The encoder outputs $64$ feature maps which is split  into $63$ public feature maps and $1$ privacy feature map. For the discriminator, it consists of four stride-2 convolution layers and two fully connected layers. For all the image obfuscation experiments, our network takes images of $128\times128\times3$ as inputs.

For toy experiment in Section \ref{Toy Experiment},  simpler models are implemented. We use two fully-connected layers for both encoder and decoder in encryption model . The number of neurons of each layers is $2-128-128$  in encoder and  $128-128-2$ in decoder. We then split the neurons of encoder into $126$  public neurons and $2$  privacy neurons. For discriminator, we use five fully-connected layers.

\subsection{Datasets}

In the following experiments, we train our model on the Large-scale CelebFaces Attributes (CelebA) dataset \cite{liu2015faceattributes}, which contains $202,599$ celebrity face images and $10,177$ identities. The images were obtained from the internet and vary extensively in terms of pose, expression, lighting, image quality, background clutter and occlusion, which is quite challenging to test the robustness of image obfuscation algorithms.

The recognition comparison experiments are conducted on Facescrub Dataset \cite{ng2014data}, CASIA-WebFace Dataset \cite{yi2014learning} and LFW dataset \cite{huang2008labeled}.

\section{Experiment}\label{experiment}

\subsection{Toy Experiment}\label{Toy Experiment}
\begin{figure}[h]
\begin{center}
   \includegraphics[width=1\linewidth]{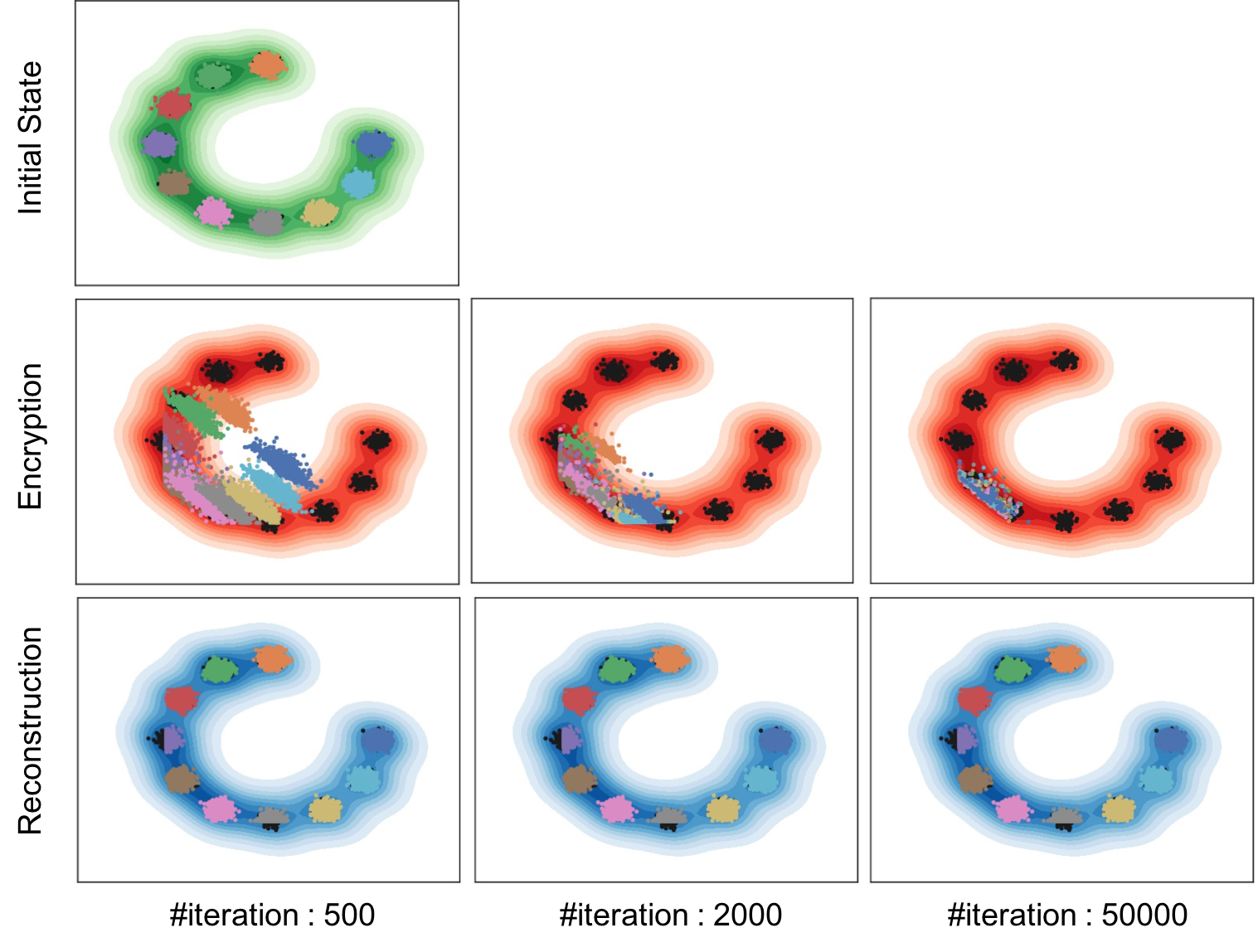}
\end{center}
   \caption{Distribution discrepancy maximization process on synthetic samples.  The distribution of ten group generated samples is distinguished by different colors. Top row: initial distribution of original samples. Medium row: the distribution of encrypted samples of different training iteration. Bottom row: the distribution of reconstructed samples of different training iteration. }
\label{fig:toy}
\end{figure}

To empirically demonstrate our explanations on distribution discrepancy maximization between two domains, we design an illustrative experiment based on 2-dimensional synthetic samples as shown in Figure \ref{fig:toy}. We generate ten group data points in 2-dimensional space distinguished by different colors. Our goal is to obfuscate these samples in the encryption domain, while maintaining them in the reconstructed results.

In Figure \ref{fig:toy}, we could intuitively observe the trend of distribution changing in the training process. Our model gradually shifts the distribution of encrypted samples away from original ten group distributions, and finally aggregates all the encrypted samples into nearly one single group. For reconstruction, the distribution of reconstructed results is highly consistent with original distribution, which supports our practice that the distribution discrepancy maximization between encrypted samples and reconstructed samples is basically equal to the maximization between the encrypted samples and the original samples. We also observe that the reconstruction converges much earlier than the encryption, and remains stable after $500$ training iterations.  Although this experiment is limited due to its simplicity, the results clearly support the validity of our proposed method.

\subsection{Comparison Results}\label{Comparison Results}

\subsubsection{Baseline}

We compare our model with three obfuscation methods, including Pixelation, Blurring, P3 \cite{ra2013p3}. For Pixelation, we downsample the images based on the scale factor x20, which could lead to better obfuscation results \cite{ryoo2017privacy-preserving}. For Blurring, we compare with Blurring radius $16$. For P3, as the smaller threshold causes the better encrypted effect, we choose the smallest threshold $1$ for comparison. In addition to existed methods, we design a MSEDNet that is a MSE based decomposition network without collaborative training. The objective function of MSEDNet is:
\begin{equation}
\begin{aligned}
\ell_{MG} = \ell_{G}^{\text{r}}-{\mathbb{E}_{x \sim {p}(x)}}[\Vert \phi(x_e) - \phi(x)\Vert_2^2].
\end{aligned}
\label{eq:MSEDNet_loss}
\end{equation}
The goal of MSEDNet is to maximize the perceptual loss between the input image and the encrypted image, and minimize the reconstruction loss between the input image and the reconstructed image.

\begin{table}[t]
\footnotesize
\setlength{\abovecaptionskip}{0.cm}
\setlength{\belowcaptionskip}{-0.cm}
\centering
\caption{Recognition accuracy of plain convolutional network on the Facescrub dataset encrypted by different methods.}
\label{table:facescrub}
\scalebox{1.0}{
\begin{tabular}{p{3cm}c}
\toprule
Method   &   Facescrub\\
\midrule
Original     & 84.6\% \\
Random & 0.19\%  \\
Pixelation(20)     & 44.4\% \\
Blurring(16)     & 41.2\% \\
P3(1)   & 23.4\%  \\
MSEDNet     &  36.1\% \\
Ours     & 3.43\% \\
\bottomrule
\end{tabular}}
\end{table}

\begin{table}[t]
\footnotesize
\setlength{\abovecaptionskip}{0.cm}
\setlength{\belowcaptionskip}{-0.cm}
\centering
\caption{Recognition accuracy of FaceNet on the Casia-WebFace \& LFW datasets encrypted by different methods.}
\label{table:casia-lfw}
\scalebox{1.0}{
\begin{tabular}{p{2cm}p{2cm}cp{2cm}c}
\toprule
Method   &   Casia-WebFace   & LFW   \\
\midrule
Original    &   87.5\%   & 98.9\%  \\
Random    &   0.0095\%   & 0.017\%  \\
Pixelation(20)   &   34.0\%   &20.9\%   \\
Blurring(16)   &   51.8\%   &54.3\%   \\
P3(1)  &   35.2\%   &21.7\%   \\
MSEDNet   &   34.7\%   &21.9\%   \\
Ours   &   0.01\%   &0.02\%   \\
\bottomrule
\end{tabular}}
\end{table}

\subsubsection{Deep Recognition Attack Model}

We follow the experimental process as proposed in \cite{mcpherson2016defeating}. We assume that one adversary can input original un-encrypted images to obfuscation algorithms (in online social network) and get the corresponding encrypted images. Therefore, we generate the training set by applying obfuscation algorithms to the original images. Then we perform supervised learning on the encrypted images to obtain the deep encrypted-image recognition models. Finally, the performance of obfuscation algorithms are measured by the accuracy of trained recognition models.

In our experiments, we first evaluate on Facescrube dataset by a plain convolutional network as the settings of \cite{mcpherson2016defeating}. Then we deploy a more powerful attack model, FaceNet \cite{schroff2015facenet}, which is a deep learning architecture consisting of convolutional layers based on GoogLeNet inspired inception models. Instead of the triplet loss presented in FaceNet, we train the attack model using softmax loss for more stable and faster convergence, which could also achieve well performed results. We choose nine-tenth encrypted face images of each celebrity in CASIA-WebFace for FaceNet model training, and evaluate on the remaining of CASIA-WebFace dataset and LFW dataset.

\subsubsection{Encryption Results Comparison}

\begin{figure}[t]
\begin{center}
   \includegraphics[width=1.0\linewidth]{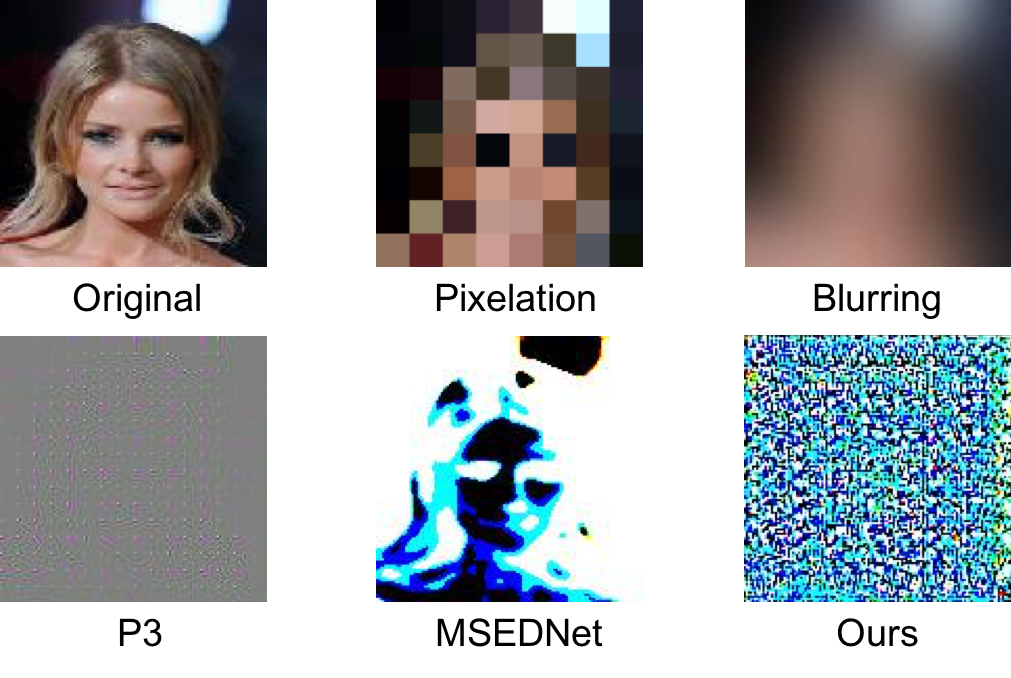}
\end{center}
   \caption{The visual encryption results comparison of different methods.}
\label{fig:qual_compare}
\end{figure}

\begin{figure}[htbp]
\begin{center}
   \includegraphics[width=0.9\linewidth]{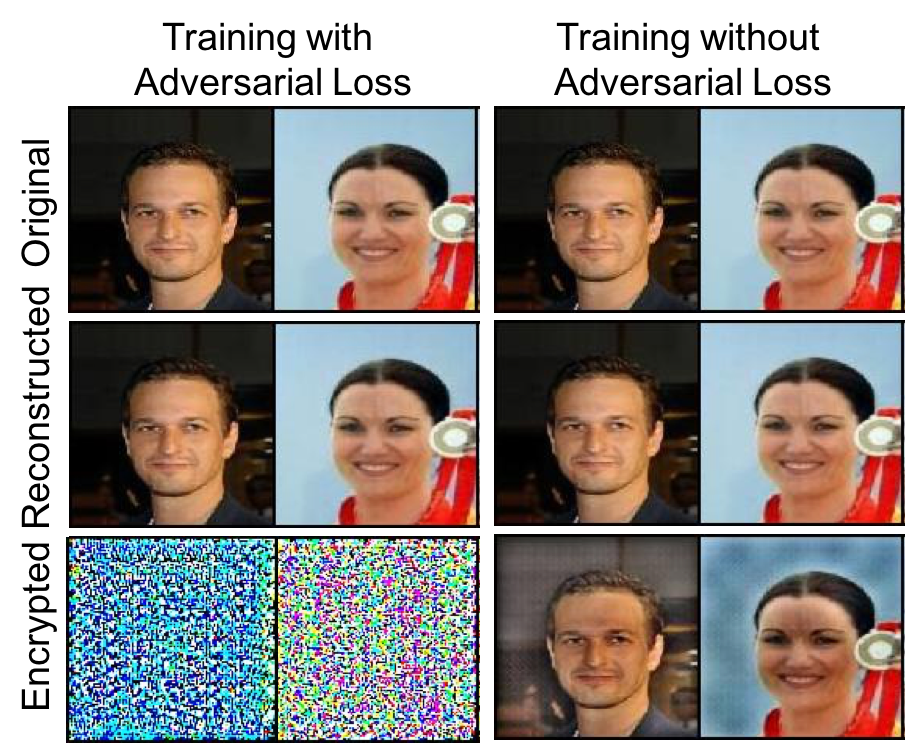}
\end{center}
   \caption{The visual results of our model with and without collaborative training scheme.}
\label{fig:eff_adv}
\end{figure}

We report the accuracy of plain convolutional neural network on Facescrub dataset in Table \ref{table:facescrub}. By applying different image obfuscation algorithms for face encryption, we can observe that the recognition accuracy of Pixelation, Blurring and P3 decreases by a large margin. However, all accuracy of these algorithms does not drop below $20\%$. Although MSEDNet tries to maximize the perceptual distance, the result also shows its insufficient ability against deep recognition model. In comparison, our method achieves $3.43\%$ accuracy, which is relatively closer to random guess. The face recognition results on the Casia-WebFace \& LFW datasets are also presented in Table \ref{table:casia-lfw}. Even a more powerful deep model FaceNet is applied for encrypted face recognition, our model still significantly outperforms other methods on these two datasets. The accuracy of the attack model on our method is also close to random guess. It is important to note that our discriminator and the deep recognition model do not have any parameter sharing and implicit relationship. Considering that the deep recognition model is a highly non-linear learning structure, it indicates that our model can produce encrypted images that are significantly different from input images through collaborative training scheme. From the visual encryption results in Figure \ref{fig:qual_compare}, we can find that the encrypted results produced by our model are also visually un-recognizable. Therefore, we verify that our model can effectively protect the sensitive information of images from the attack of deep recognition models and human visual recognition.

\begin{figure*}[t]
\begin{center}
   \includegraphics[width=1.0\linewidth]{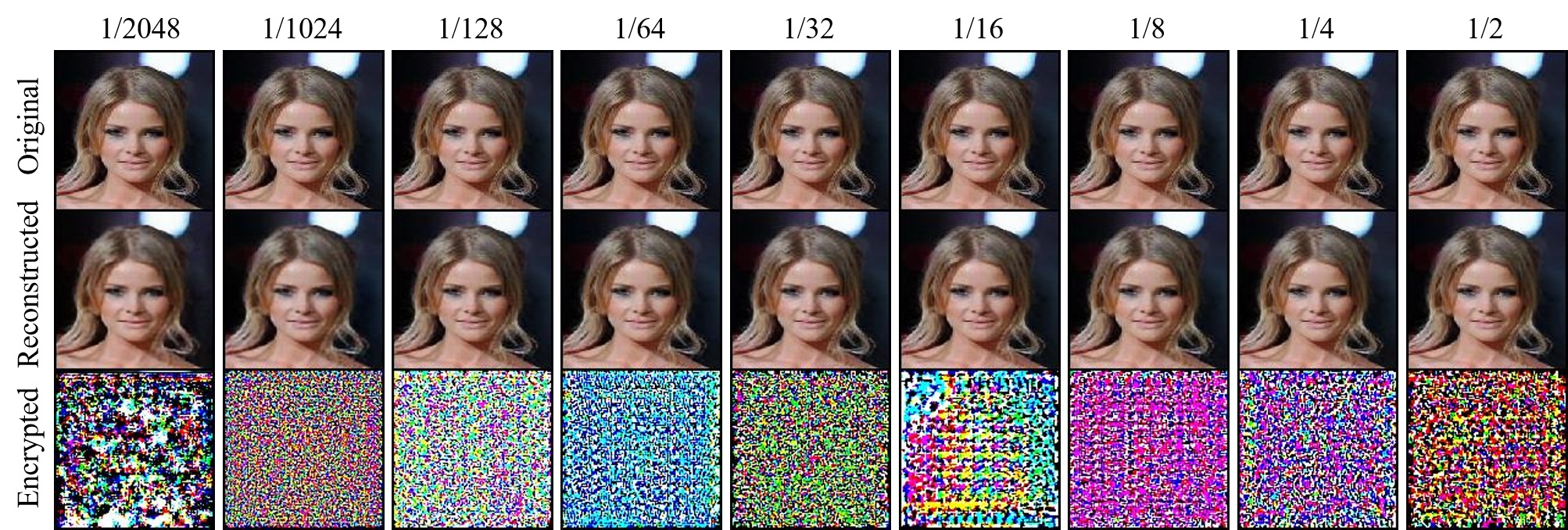}
\end{center}
   \caption{The visual results of different proportions of privacy part.}
\label{fig:ratio}
\end{figure*}

\subsubsection{Additional Comparison with Adversarial Perturbation Methods}
In this subsection, we compare with two adversarial perturbation methods, including Fast Gradient Step Method (FGSM) \cite{goodfellow2015explaining} and Universal Adversarial Perturbations (UAP) \cite{mopuri-bmvc-2017}, which aims to clarify the difference between adversarial perturbation methods and our model. We evaluate the face recognition accuracy on LFW dataset with FaceNet model pre-trained on original images of Casia-WebFace Dataset, since adversarial perturbation aims to make a existing deep model misclassify images of perturbations. As shown in Table \ref{table:adversarial_attackj}, our model can greatly mislead the pre-trained FaceNet compared with adversarial perturbation methods, which indicates that images encrypted by removing privacy feature in the latent space ensures much better privacy preserving than adversarial perturbation.

\begin{table}[t]
\footnotesize
\setlength{\abovecaptionskip}{0.cm}
\setlength{\belowcaptionskip}{-0.cm}
\centering
\caption{Recognition accuracy comparison between adversarial perturbation methods and our model with  FaceNet trained on original Casia-WebFace dataset.}
\label{table:adversarial_attackj}
\scalebox{1.0}{
\begin{tabular}{p{5cm}c}
\toprule
Methods   &   Accuracy   \\
\midrule
Origin   &   98.9\%  \\
UAP \cite{mopuri-bmvc-2017}     &  63.0\%  \\
FGSM \cite{goodfellow2015explaining} &   19.8\%  \\
Ours &   0.01\%  \\
\bottomrule
\end{tabular}}
\end{table}

\subsection{Effectiveness of Collaborative Training Scheme}
In this section, we analyze the effectiveness of collaborative training scheme. We compare the results of our model and model without collaborative training scheme. The visual results are shown in Figure \ref{fig:eff_adv}. In addition to the recognition accuracy of FaceNet on the Casia-WebFace encrypted by our model, to show the quality of reconstructed image and compare the quality degradation between reconstructed image and encrypted image, PNSR results of reconstructed and encrypted images are also provided in Table \ref{table:eff_adv}. We can observe that, by directly removing feature maps without proposed collaborative training scheme, the encrypted image maintains most of recognizable information of the input image and can not guarantee the privacy safety. In addition, the model with collaborative training can reconstruct image as well as the model without collaborative training, which indicates that the collaborative training pushes privacy information into the privacy part without losing the overall image information. Therefore, we have shown that collaborative loss is essential for excluding privacy information from the public part and producing highly unrecognizable encrypted images.
\begin{table}[t]
\footnotesize
\setlength{\abovecaptionskip}{0.cm}
\setlength{\belowcaptionskip}{-0.cm}
\centering
\caption{PSNR (dB) of reconstructed and encrypted images, and defense performance of our model with/without collaborative training scheme.}
\label{table:eff_adv}
\scalebox{1.0}{
\begin{tabular}{c c c c}
\toprule
Collaborative Loss   &   Reconstructed   & Encrypted  & Accuracy \\
     &  (dB)& (dB) &\\
\midrule
With & 33.31     &  4.65   &  0.01\%\\
Without & 33.65     &  17.06    &  79.6\% \\
\bottomrule
\end{tabular}}
\end{table}

\subsection{Robustness to the Proportion of Privacy Part}\label{privacy_proportion}
We have shown that our model can achieve extraordinary encryption performance with one $64$th of deep features extracted as privacy part, we here continue to explore our model's robustness to the different proportions of privacy part, including $1/2048$, $1/1024$, $1/128$, $1/64$, $1/32$, $1/16$, $1/8$, $1/4$ and $1/2$. The visual results of models with different proportions of privacy part are shown in Figure \ref{fig:ratio}. Similarly, we show the reconstruction quality, encryption quality and accuracy of different proportions compared with P3 in Table \ref{table:ratio}. From the experiment results, we can see that our model is quite robust to the various proportions of privacy part in terms of reconstruction quality and encryption accuracy. In addition, our model can achieve comparable reconstruction quality compared with P3 that are based on JEPG coding standard, while much lower proportion of privacy part than p3 is required by our model. We choose $1/64$ as our main configuration to achieve a trade-off among reconstruction quality, encryption accuracy and proportion.

\begin{table}[t]
\footnotesize
\setlength{\abovecaptionskip}{0.cm}
\setlength{\belowcaptionskip}{-0.cm}
\centering
\caption{PSNR (dB) results and defense performance of different proportions of privacy part .}
\label{table:ratio}
\scalebox{0.89}{
\begin{tabular}{c c c c c c}
\toprule
Method  & Proportion   &   Reconstructed     & Encrypted  & Accuracy \\
&&(dB)&(dB)&\\
\midrule
P3(20) & 19.68\%($\approx1/5$)    & 37.86    &  12.10   &  67.2\% \\
P3(10) & 23.5\%($\approx1/4$)    &   35.03     &  12.00  & 62.3\% \\
P3(1) &  55.62\%($\approx1/2$)  &   30.83   &  11.85  & 35.2\%  \\
\hline
\multirow{7}*{Ours}    &1/2048 & 30.12    &  5.75  & 0.055\%\\
~&1/1024 & 30.42     &  5.19  & 0.052\% \\
~&1/128 & 33.31     &  4.65   & 0.016\%\\
~&1/64 & 34.00     &  4.41   & 0.01\%\\
~&1/32 & 33.99     &  5.17   & 0.01\%\\
~&1/16 & 33.96     &  4.82   & 0.01\%\\
~&1/8 & 33.99     &  4.75   & 0.01\%\\
~&1/4 & 34.54    &  4.64   & 0.01\%\\
~&1/2 & 34.83     &  5.02   & 0.01\%\\
\bottomrule
\end{tabular}}
\end{table}

\section{Conclusion}\label{conclusion}

In this paper, we propose to maximize the distribution discrepancy for image privacy preserving. Given an input image, our model decomposes it into public feature and privacy feature, and generates a reconstructed image and a encrypted image accordingly. To produce distribution discrepancy between the input image and the encrypted image, we introduce a collaborative training scheme, where a discriminator and a encryption model are trained to optimize the same objective. We theoretically prove that the collaborative training scheme maximizes the distribution discrepancy. We conduct sufficient experiments to validate effectiveness of our proposed technique. Compared with existing image obfuscation methods, our model can produce satisfactory defense under the attack of deep recognition model while maintaining the quality of reconstruction.

\clearpage
\bibliographystyle{aaai}
\bibliography{bibliography}

\end{document}